\documentclass[conference]{IEEEtran}

\usepackage{setspace}
\usepackage{graphicx,url}
\usepackage[T1]{fontenc}
\usepackage{soulutf8}
\usepackage{xcolor}
\usepackage{booktabs}
\usepackage{hyperref}
\hypersetup{hidelinks}
\usepackage{multicol}
\usepackage{multirow}
\usepackage{epsfig}
\usepackage{subcaption}
\usepackage{amsfonts}
\usepackage{mathtools}
\usepackage{amsmath}
\usepackage{graphicx}
\usepackage{multirow}
\usepackage{multicol}
\usepackage{soulutf8}

\usepackage{url}
\usepackage{algorithm}
\usepackage{algpseudocode}
\PassOptionsToPackage{noend}{algpseudocode}

\errorcontextlines\maxdimen

\makeatletter
\newcommand*{\algrule}[1][\algorithmicindent]{\makebox[#1][l]{\hspace*{.5em}\vrule height .75\baselineskip depth .25\baselineskip}}%

\newcount\ALG@printindent@tempcnta
\def\ALG@printindent{%
    \ifnum \theALG@nested>0
        \ifx\ALG@text\ALG@x@notext
            \addvspace{-3pt}
        \else
            \unskip
            \ALG@printindent@tempcnta=1
            \loop
                \algrule[\csname ALG@ind@\the\ALG@printindent@tempcnta\endcsname]%
                \advance \ALG@printindent@tempcnta 1
            \ifnum \ALG@printindent@tempcnta<\numexpr\theALG@nested+1\relax
            \repeat
        \fi
    \fi
    }%
\usepackage{etoolbox}
\patchcmd{\ALG@doentity}{\noindent\hskip\ALG@tlm}{\ALG@printindent}{}{\errmessage{failed to patch}}
\makeatother

\title{A Clustering-Based Method for Automatic Educational Video Recommendation Using Deep Face-Features of Lecturers}
\author{\IEEEauthorblockN{Paulo R. C. Mendes, Eduardo S. Vieira, Álan L. V. Guedes, Antonio J. G. Busson, and Sérgio Colcher}
\IEEEauthorblockA{TeleMidia Lab - Department of Informatics\\Pontifical Catholic University of Rio de Janeiro\\Rio de Janeiro, Brazil\\\{paulo.mendes, eduardo, alan, busson\}@telemidia.puc-rio.br, colcher@inf.puc-rio.br}}

\begin{document}

\maketitle

\begin{abstract}
Discovering and accessing specific content within educational video bases is a challenging task, mainly because of the abundance of video content and its diversity. Recommender systems are often used to enhance the ability to find and select content. But, recommendation mechanisms, especially those based on textual information, exhibit some limitations, such as being error-prone to manually created keywords or due to imprecise speech recognition.
This paper presents a method for generating educational video recommendation using deep face-features of lecturers without identifying them.
More precisely, we use an unsupervised face clustering mechanism to create relations among the videos based on the lecturer's presence.
Then, for a selected educational video taken as a reference, we recommend the ones where the presence of the same lecturers is detected.
Moreover, we rank these recommended videos based on the amount of time the referenced lecturers were present. 
For this task, we achieved a mAP value of 99.165\%.

\end{abstract}


\section{Introduction}\label{sec:intro}

The traditional paradigm of classroom courses centered on the physical presence of a teacher has been gradually giving space to online and hybrid courses, which enables the emergence of VTEs (Virtual Teaching Environment) and MOOCs~(\textit{Massive Open Online Courses}, such as Udacity,\footnote{\url{https://udacity.com}} Coursera,\footnote{\url{https://coursera.org}} and EdX).\footnote{\url{https://edx.org}}
For example, in 2018, a study~\cite{pearson2018} has shown that almost 59\% of people aged 14 to 23 prefer YouTube as a learning tool rather than printed books, with 55\% of them also saying that YouTube has contributed to their education.
Recently, due to the covid-19 outbreak, the world has experienced an unprecedented usage of virtual education~\cite{sun2020coronavirus}, and some say that this model of education came to stay.\footnote{https://www.cnbc.com/2020/05/20/post-pandemic-remote-learning-could-be-here-to-stay.html}

If, on the one hand, the abundance of educational videos can contribute to and facilitate learning, on the other hand, it also makes it challenging to discover and access the content of interest~\cite{dias2017approach}.
This issue is usually addressed by a proactive user search (using queries, for example), or by automatic recommendations made by specialized systems.

Recommendation mechanisms are usually based on two methods: \textit{collaborative filtering} and \textit{content-based filtering}. 
In collaborative filtering, the system groups users based on their common interest on items, using users' preferences, rates, purchases or accesses to those items. With this approach, 
knowledge about the item's content is not needed; the recommendation is purely based on the relationship between users and items.  The content-based filtering, differently, requires items' description; similar items are the ones recommended to the user.

In general, the current video recommendation methods are heavily dependent on textual information from the video, such as labels (\textit{i.e.} keywords)~\cite{mahajan2015optimising,omisore2014personalized}, or automatically generated captions \cite{barrere2020utilizaccao} from the lecturer speech.
These systems face problems such as errors introduced by manually inserted labels and by imprecise speech recognition.
In our research, we aim to investigate methods that are able to perform video recommendations that are not based on content nor on any error-prone textual descriptions, but solely on lecturers' presence. Notice that this approach does not necessarily have to completely replace textual-based recommendations; in fact, it can be easily used as an additional aid to enhance the ability to find content in any system.

Face detection methods have been attracting the attention of researchers for more than two decades~\cite{survey66}. 
Nowadays, it is used for surveillance, video analytics systems, smart shopping, automatic face tagging in photo collections, investigative tools that search for identities in social networks based on face images, and thousands of other applications in our daily lives.
For instance, Facer~\cite{hazelwood2018applied} is the Facebook's face detection and recognition framework; given a photograph, it first detects all the faces, and then runs a deep model to determine the likelihood of that face belonging to one of the top-N user friends.
This allows Facebook to suggest which friends the user might want to tag within the uploaded photographs. 

This work aims at recommending educational video content based on lecturers' presence.
To do that, we take advantage of face detection methods.
More precisely, we detect lecturers in a video taken as a reference and perform a clustering based on the face of these lecturers in different videos.
Given these clusters, we extract their \textit{centroids} (explained in Section~\ref{sec:method}), and perform another clustering step for creating a relationship between videos that share the presence of the same lecturers.
Finally, we rank the recommended videos based on the amount of time the referenced lecturers were present.
A particular feature of this approach is that it can be done without supervision, allowing for new videos to be automatically analyzed.
Moreover, our approach permits the creation of timelines based on lecturers' presence that 
can be used in the search for specific parts of a content where only specific lecturers' are present.
To evaluate our recommendation ranking, we use the mAP~(Mean Average Precision) metric, which is commonly used in information retrieval evaluation tasks~\cite{manning2009introduction}.

The remainder of this paper is structured as follows.
Section~\ref{sec:related_work} discusses some related work.
Then, we present our method in Section~\ref{sec:method}.
Section~\ref{sec:dataset} presents the used dataset, followed by 
Section~\ref{sec:experiments}, that shows the experiments to validate the face clustering and the video recommendation ranking mechanisms.
Finally, Section~\ref{sec:remarks} brings our final remarks.

\section{Related Work}
\label{sec:related_work}

We have organized the related work into two groups.
In the first, we grouped works that share our goal of educational video recommendation but do not necessarily use face-embeddings~(deep face-features).
The second group is the one in which every work addresses the task of face recognition in videos.

Regarding \textit{Educational Video Recommendation}, we cite works based on content-filtering.
These works perform analyses and comparisons using the video textual description or speech recognition performed on them. 
Omisore \textit{et. al.}
\cite{omisore2014personalized}, for example, propose combining \textit{fuzzy} techniques to recommend books with content suitable for students based on their reading histories in a digital library, while Mahajan \textit{et. al.}
\cite{mahajan2015optimising} propose, given a reference video,  mining social media, and web for suggesting links for a student to visit.
Moreover, 
Barrére \textit{et. al.}
~\cite{barrere2020utilizaccao} use texts from speech recognition to create recommendations.
These works are only based on textual characteristics~(or content converted to it) for performing recommendations.
Our work focuses on using a visual part of the video, more precisely the presence of lecturers.

Works based on \textit{Video Face Recognition} usually apply deep learning models for the task. DeepFace \cite{taigman2014deepface} and DeepID \cite{sun2014deep}, for example, use a CNN (Convolutional Neural Network) with a fully-connected layer output to produce a representation of high-level features (face embeddings) from an input image, followed by a softmax layer to indicate the identity of classes. 
Other approaches, such as FaceNet \cite{schroff2015facenet}, can directly measure the similarity among faces using euclidean space. 
Yang \textit{et al.}~\cite{yang2017neural} proposed a deep network for video face recognition called NAN (Neural Aggregation Network). They use a CNN to generate the embeddings, followed by an aggregation module that consists of two attention blocks which adaptively aggregate the feature vectors to form a single feature inside the convex hull spanned by them. 
Rao \textit{et al.} \cite{rao2017attention} proposed a method for video face recognition based on attention-aware deep reinforcement learning. They formulated the process of finding the attention of videos as a Markov decision process and training the attention model without using extra labels. Unlike existing attention models, their method takes information from both the image space and the feature space as the input to make use of face information that is discarded in the feature learning process. 
Sohn \textit{et al.} \cite{sohn2017unsupervised} proposed an adaptative deep learning framework for image-based face recognition and video-based face recognition. Given an embedding generated by a CNN, their framework adaptation is achieved by (1) distilling knowledge from the network to a video adaptation network through feature matching, (2) performing feature restoration through synthetic data augmentation, and (3) learning a domain-invariant feature through an adversarial domain discriminator.

Like \cite{yang2017neural, rao2017attention, sohn2017unsupervised}, our method uses a CNN to generate face embeddings from face images, with the difference that we use an unsupervised cluster-based method to compare the similarity among faces extracted from videos.

\section{Method}
\label{sec:method}

Our method intends to recommend educational videos based on the lecturers that appear in each video, so that, when a person watches a video, other videos containing the same lecturers are recommended.
For didactic purposes we decided to divide our exposition in two phases: (i)~\emph{video representation} and (ii)~\emph{video recommendation}, which are described in Sections~\ref{subsec:video_representation} and \ref{subsec:video_recommendation} respectively.

\subsection{Video Representation}
\label{subsec:video_representation}

\begin{figure*}[!ht]
  \centering
  \includegraphics[width=\textwidth]{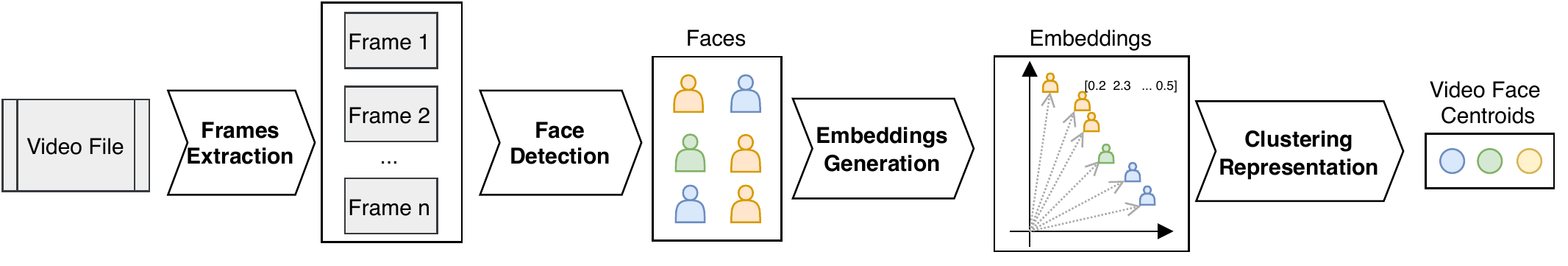}
  \caption{Lecturers representation process in video. This process receives a video file and returns the centroids of the clusters that ideally represent each of the lecturers present in the video file.}
  \label{fig:video_clustering}
\end{figure*}

The objective of this phase is to represent each video with vectors~(centroids) of the lecturers that appear on it. 
Fig. \ref{fig:video_clustering} shows the pipeline we propose for this phase, described in the remainder of this subsection.
It is divided into four steps: \emph{Frames Extraction}, \emph{Face Detection}, \emph{Embeddings Generation} and \emph{Clustering Representation}.

First, we perform the \textit{Frames Extraction} by receiving a video file as input and extracting its frames according to a given frame rate. 
Next, for each of the frames, the \textit{Face Detection} step uses an object detection model for detecting faces in each of them. 
The face detection model is responsible for returning the bounding boxes of the faces present in the image giving the $x$ and $y$ axes coordinates of the upper-left corner and of the lower-right corner of the rectangle that establishes the visual limits encapsulating each face. 
With these bounding boxes, we can isolate and extract the bounded images, obtaining a dataset composed of images of faces only.

The objective of the \textit{Embeddings Generation} step is to represent each face image as a vector in $\mathbb{R}^{n}$.
To achieve that, it processes each of the faces generated in the previous step through a CNN that generates their embeddings.
An embedding is a representation of the input in a lower dimensionality space.
Ideally, an embedding captures some semantics of the input, e.g. by placing semantically similar inputs close together in an embedding space.
Therefore, at the end of this step, we have all faces represented as embeddings.

In the \textit{Clustering Representation} step, we group embeddings~($e$), and, consequently, faces that are close enough in the embedding space using a clustering algorithm.
Clustering is the task of dividing a set of data points, embeddings in this case, into a number of groups~(called \emph{clusters}) such that data points in a given group are similar to other data points in the same group and dissimilar to the data points in other groups.
The clustering process should produce a partition of the faces present in the frames, hopefully with each generated cluster representing a specific person; moreover, the union of all clusters covers the whole set of faces found in the video. 

As most of the clustering algorithms require the number of clusters as parameter, we use a strategy~(defined in Algorithm \ref{clustering_alg}) based on the \emph{Silhouette Score} ($s$) \cite{rousseeuw1987silhouettes}, that corresponds to the mean of the \emph{Silhouette Coefficient} ($\sigma$) of all samples. This coefficient for each sample is 
\begin{equation}
  \label{equation:silhouette}
  \sigma = \frac{b-a}{max(a,b)}
\end{equation}
where $a$ is the mean distance from a sample to all other samples in the same cluster, and $b$ is the mean distance from a sample to all other samples in the closest cluster to that sample.
In this way, the best value is 1 and the worst is -1. Values close to 0 indicate overlapping clusters, whereas negative values usually indicate that a sample has been assigned to the wrong cluster since a different cluster is more similar.

\begin{algorithm}
\caption{Iteratively finding the best clustering configuration for unknown number of clusters.}\label{clustering_alg}
\begin{algorithmic}[1]
\Procedure{BlindClustering}{$e, t,\omega$}\
\State $n_K\gets 1$
\State $s_{max}\gets -1$
\State $t_{cur}\gets 0$ 

\While{$t_{cur} \leq t \And n_K < |e|$}
    \State $n_K\gets n_K+1$
    \State $K_{cur}\gets Clustering(e, n_K)$
    \State $s \gets SilhouetteScore(K_{cur})$
    \If{$s < s_{max}$}
        \State $t_{cur}\gets t_{cur}+1$
    \Else
        \State $K\gets K_{cur}$
        \State $t_{cur}\gets 0$
        
        \If{$s > s_{max}$}
            \State $s_{max} \gets s$
        \EndIf
    \EndIf
\EndWhile
\If{$s_{max} < \omega$}
    \State $K\gets OneCluster(e)$
\EndIf
\State \textbf{return} $K$
\EndProcedure
\end{algorithmic}
\end{algorithm}

With the strategy defined in Algorithm \ref{clustering_alg}, we increase the number of clusters until the maximum Silhouette Score decreases to more than $t$ times in a row or until it reaches the maximum number of clusters~(lines 5-18), which is the number of embeddings~($|e|$).
The \texttt{Clustering} procedure (line 7) can be substituted by any clustering algorithm that requires the number of clusters in advance.
When the iteration stops, we return the clustering configuration with the highest Silhouette Score.
Since the Silhouette Coefficient requires at least two clusters, it would not be possible to compute the Silhouette Score for a clustering configuration with only one cluster~(there are only faces of a single person).
To overcome this problem, we start with 2 clusters consecutively increasing it as described above. Then, if the returned clustering configuration has a Silhouette Score smaller than a threshold $\omega$, that probably indicates overlapping, we say that all faces belong to one single cluster~(lines 19-20).

Next, we compute the clusters' centroids for each of the clusters $k~\in~K$ where $K$ is the best clustering configuration found with the \emph{Silhouette Score}.
A centroid $c_k$ for each cluster is the mean of the elements present in the cluster, and can be defined as follows
\begin{equation}
  \label{equation:centroid}
  c_k = \sum_{a~\in~k}{\frac{a}{|k|}}
\end{equation}
where $a$ represent each element of a cluster $k$.

By the end of this phase, we have each video in the dataset represented by its centroids where, ideally, each centroid represents a lecturer present in the video.
We also record the frames where each lecturer is present.

\subsection{Video Recommendation}
\label{subsec:video_recommendation}

\begin{figure*}[!ht]
  \centering
  \includegraphics[width=0.8\textwidth]{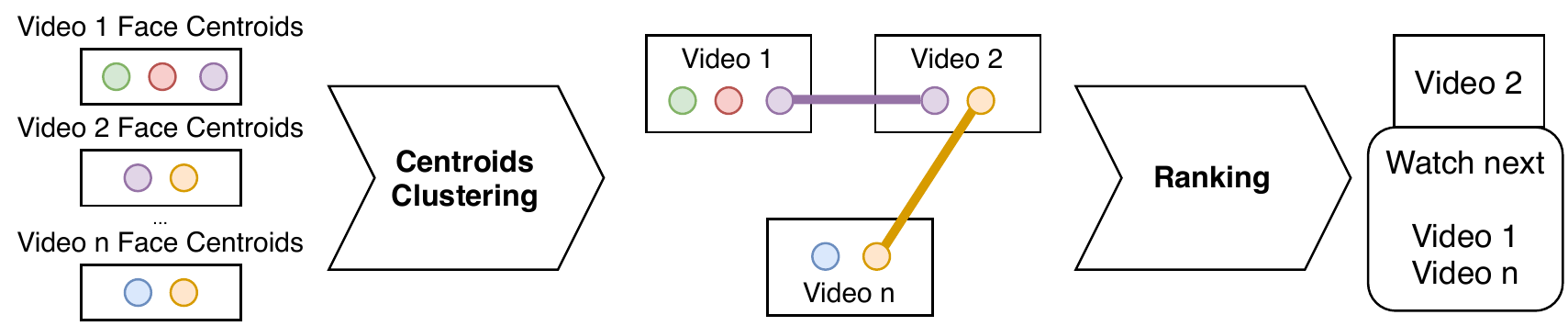}
  \caption{Video Recommendation based on Lecturers Centroids Clustering. This pipeline receives the centroids of lecturers from all the videos in the dataset, then it creates relationships among videos that share the presence of the same lecturers. Finally, it performs ranking of recommended videos for each of the videos in the dataset. This ranking is based on the number of lecturers shared and their time presence.}
  \label{fig:video_recommendation}
\end{figure*}

This phase aims at recommending videos by the lecturers present in it and in the other videos.
It is divided in two steps: \emph{Centroids Clustering} and \emph{Ranking}, as depicted in Fig. \ref{fig:video_recommendation}.

First, we gather the centroids from the videos of the dataset as one single set and perform the \textit{Centroids Clustering}.
For performing this clustering, we also use the strategy for an unknown number of clusters described in Algorithm \ref{clustering_alg}.
By doing that, we group centroids from the same lecturer that are in different videos. For instance, in Fig. \ref{fig:video_recommendation}, one can see that the \emph{purple lecturer} is present in both Videos 1 and 2, while the \emph{orange lecturer} is present in both Videos 2 and n. By the end of this step, we have the group $L$ of lecturers present in the dataset of videos $V$, and we can also denote $L_v$ as the group of lecturers present in video $v$.

Next, based on these relationships among different videos, we perform \textit{Ranking}, by recommending videos in which lecturers of the current video are present. 
For doing that, we compute a similarity score using the presence of the lecturers in the current video and the presence of these same lecturers in the other video.
Let $p_{l,v}$ denote the percentage of frames in which the lecturer $l \in L_v$ is present in video $v \in V$. For each video $v \in V$ and $u \in V-v$ we compute a score of similarity $S_{v,u}$.

\begin{equation}
  S_{v,u} = \sum_{l~\in~L_v}{p_{l,v}\cdot{p_{l,u}}}
\end{equation}

Finally, using this score, for each video $v$ we compute a ranking $R_{v}$ where $R_{v,i}$ denotes the \emph{i-greatest} $S_v$ and $R_{v,i}\ge~R_{v,i+1}$ for all $i~\in~1...n_v$, where $n_v$ is the number of videos $u$ in which $S_{v,u}>~0$. 
In this way, the more lecturers a video have in common with the reference video, and the more time these lecturers are present in both videos, the higher the video is positioned in the ranking of the reference video.  

By the end of this phase, we have a ranking of recommended videos for each video in the dataset.
It is important to notice that our method is unsupervised and does not require the information of the lecturers in advance.
Consequently, we do not store any information regarding the identity of the lecturers, respecting their privacy.

\section{Dataset}
\label{sec:dataset}
The experiments were conducted using a dataset created in the context of this work.
It is composed of 98 educational videos publicly available on YouTube.\footnote{\url{https://www.youtube.com/channel/UCT0JugAtGmqiYkwxFZOwAtg}}$^{,}$\footnote{\url{https://www.youtube.com/user/deboraaladim}}

Each video contains at least one lecturer; moreover, some videos could have some special participation or collaboration. Thus, each video is annotated to contain between 1 to 5 people. In total, 16 people are present in the dataset. Each person has an average presence of 6.67\% in the videos, and their identities are known and ready to be used to assist in the nominal dataset organization.

Regarding the duration, the videos vary between 00m:30s and 1h:49m:01s. 
The average duration of the videos is 23m:34s, with a standard deviation of 23m:05s. 
The high value of the standard deviation for the time estimates indicates that the videos are not in the same time range, and therefore have a wide duration variety. 

\section{Experiments}
\label{sec:experiments}
First we compute the centroids that represent each lecturer in each video using the process described in Section \ref{subsec:video_representation}. Next we perform the video recommendation using the process described in Section \ref{subsec:video_recommendation}.

For representing the video files in the dataset, we start by performing \emph{Frames Extraction} for each video file using a frame rate of 1 frame per second~(fps). 
Next, in the \emph{Face Detection} step, we use MTCNN \cite{mtcnn} (Multitask
Cascaded Convolutional Networks), which is widely used for the face detection task~\cite{mtcnn1, mtcnn2, mtcnn3}. 
Once we have detected the faces of lecturers in the video frames, we perform \emph{Embeddings Generation} using SE-ResNet-50 \cite{senet}~(SeNet-50 for short) that generates embeddings on the $\mathbb{R}^{2048}$ feature space. 
We used the architecture and weights pre-trained on the VGGFace2 dataset \cite{cao2018vggface2}, available on the \emph{keras-vggface} library.\footnote{\url{https://github.com/rcmalli/keras-vggface}} 
The VGGFace2 dataset contains 3.31 million images of 9,131 subjects and has large variations in pose, age, illumination, ethnicity, and profession. 
Finally, we use Algorithm \ref{clustering_alg} in the \emph{Clustering Representation} step with the parameters $t=5$, $\omega=0.2$, and the Ward Agglomerative Clustering~\cite{ward1963hierarchical} as the \texttt{Clustering} procedure using its implementation in the \emph{scikit-learn}~\cite{scikit-learn} library. The Ward Agglomerative Clustering algorithm merges pairs of clusters that minimize the \emph{ward} criterion, which is the variance of the clustering being merged.
By using this method, at each step, the algorithm finds the pair of clusters that lead to a minimum increase in total within-cluster variance after merging. Finally, we compute the centroids of each of the clusters generated.

For performing the video recommendation task, we gather the centroids~(that represent each lecturer in the video) from all videos in the dataset. Next, we perform the process described in Section \ref{subsec:video_recommendation}. For \emph{Centroids Clustering}, we also use Algorithm \ref{clustering_alg} with the same parameters $t=5$, $\omega=0.2$, and the Ward Aglommerative Clustering as \texttt{Clustering} procedure. Finally, based on the clusters generated, we perform the \emph{Ranking} step. 

The remainder of this Section describes the evaluation of the centroids clustering (Section~\ref{subsec:clustering_evaluation}) and the evaluation of the video recommendation (Section~\ref{subsec:recommendation_evaluation}).

\subsection{Centroids Clustering Evaluation}
\label{subsec:clustering_evaluation}

\begin{figure}[!ht]
  \centering
  \includegraphics[width=0.9\linewidth]{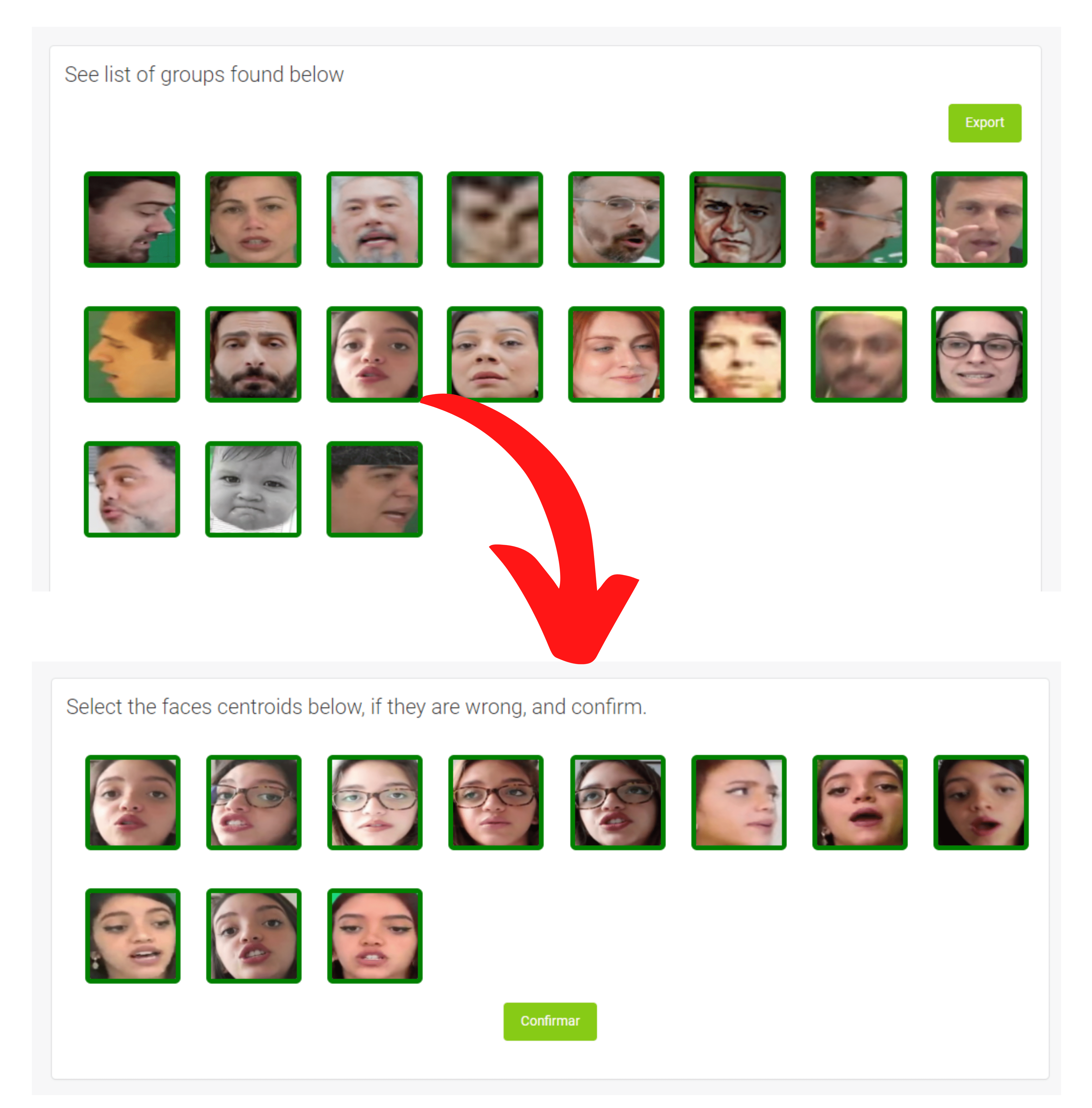}
  \caption{Centroids images correction in \textit{VideoFacesTool}. On the top, each image represents one lecturer. When a lecturer is selected, the tool displays all appearances~(centroids) of that lecturer in different videos. The user can then mark each of these appearances as correct or wrong.}
  \label{fig:TelaDeGrupos}
  
\end{figure}

Our evaluation aims at discovering the precision achieved by the clustering over the centroids.
More precisely, we want to evaluate how well our approach identified that the same lecturer is present in different video files. For this task, we require some human feedback. 
To receive that feedback, we developed an application, called \textit{VideoFacesTool} consisting of a graphical web interface. 
The tool allows participants to import a file, which contains information after the \emph{Centroids Clustering} step, so that we have the set of centroids and a sample face image of it. 
Inside the tool, a cluster is called a \emph{group}.
After being imported, faces are visually organized according to the group to which they belong and when a group is selected, all face centroids from that group are displayed, as shown in Fig. ~\ref{fig:TelaDeGrupos}.

Each face centroid has a Boolean property, which indicates whether it is correctly grouped~(it belongs to a group in which all the face centroids are from the same lecturer) or not. 
If there is an error, the participant can indicate it. 

\begin{figure}[ht]
  \centering
  \includegraphics[width=0.4\textwidth]{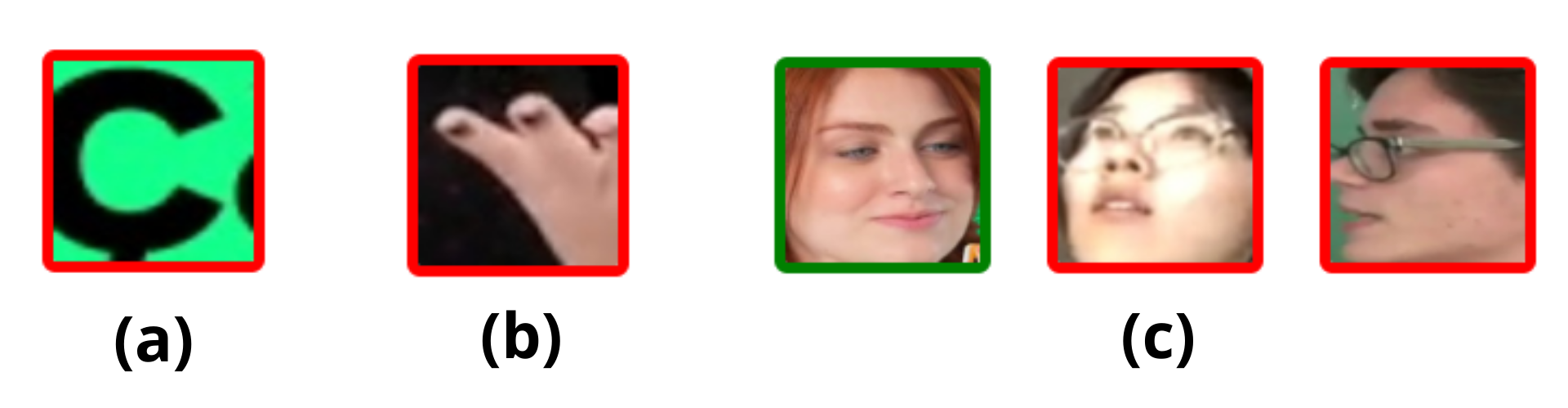}
  \caption{Examples of wrong faces centroids. (a) a part of an icon (b) a hand and (c) face centroids that are not from the same lecturer}.
  \label{fig:TelaDeErros}

\end{figure}

A total of 5 participants collaborated in the evaluation session. They were advised to mark a face centroid as wrong if it represents
(a) an object, or 
(b) a part of the human body, or  
(c) a lecturer other than the lecturer in the group. 
Fig.~\ref{fig:TelaDeErros} shows examples of these types of errors, and Table~\ref{TabelaResultadosVideoAnnotationTool} provides an overview of the evaluations obtained from each participant.
It is important to notice that these results do not reflect the recommendation of educational videos, they only evaluate the \emph{Centroids Clustering} step.
For instance, we could have a group of face centroids of people that appear for a few amount of time in the videos and are not lecturers.
This case, of course, would reduce the precision of the \emph{Centroids Clustering} step. 
However, our method for video recommendation and ranking is robust to these cases, as it considers the amount of time that a person appears for scoring the recommended videos.
With the evaluation completed, it is possible to export the analysis information with the number of right and wrong face centroids.

\begin{table}[!ht]
\caption{Results of visual evaluation of faces centroids showing the number of correct and wrong centroids classified by each participant.}

\begin{tabular}{cccc}\hline
  Participant & \#Correct & \#Wrong & Precision \\\cline{1-4}
  p1 & 163 & 62 & 72.44\% \\
  p2 & 160 & 65 & 71.11\% \\
  p3 & 164 & 61 & 72.89\% \\
  p4 & 164 & 61 & 72.89\% \\
  p5 & 162 & 63 & 72.00\% \\
  Avg & 162.6 & 62.4 & 72.266\% \\
\end{tabular}
\centering
\label{TabelaResultadosVideoAnnotationTool}
\end{table}

\subsection{Recommendation Evaluation}
\label{subsec:recommendation_evaluation}
We evaluate our approach based on the relevance of the videos recommended. A video is considered relevant to another if they have at least one lecturer in common.
To verify that, we use the information of the lecturers' presence available on our dataset.

\begin{table*}[!ht]
\small
\centering
\caption{Results obtained with our approach with different thresholds of time presence for a lecturer to be considered as present in a video.}
\label{tab:results}
\begin{tabular}{ccccccccc}
\hline
\textbf{Thershold} & \textbf{MeanR} & \textbf{MinR} & \textbf{MeanP} & \textbf{MinP} & \textbf{MeanF1} & \textbf{MinF1} & \textbf{mAP} & \textbf{MinAP} \\ \hline
0\%                & 0,88851        & 0,45455       & 0,64681        & 0,20370       & 0,70971         & 0,33333        & 0,98641      & 0,58597        \\
1\%                & 0,88851        & 0,45455       & 0,64681        & 0,20370       & 0,70971         & 0,33333        & 0,98641      & 0,58597        \\
2\%                & 0,88851        & 0,45455       & 0,64885        & 0,20370       & 0,71171         & 0,33333        & 0,98641      & 0,58597        \\
3\%                & 0,88851        & 0,45455       & 0,67368        & 0,21569       & 0,73086         & 0,34921        & 0,98642      & 0,58597        \\
4\%                & 0,88851        & 0,45455       & 0,67368        & 0,21569       & 0,73086         & 0,34921        & 0,98642      & 0,58597        \\
5\%                & 0,88851        & 0,45455       & 0,69923        & 0,22449       & 0,74930         & 0,36066        & 0,98642      & 0,58597        \\
6\%                & 0,88851        & 0,45455       & 0,73615        & 0,25714       & 0,77648         & 0,40000        & 0,98642      & 0,58597        \\
7\%                & 0,88742        & 0,45455       & 0,77849        & 0,31429       & 0,80768         & 0,45833        & 0,98642      & 0,58597        \\
8\%                & 0,88742        & 0,45455       & 0,78408        & 0,31429       & 0,81111         & 0,45833        & 0,98643      & 0,58597        \\
9\%                & 0,88742        & 0,45455       & 0,80171        & 0,33333       & 0,82410         & 0,47826        & 0,98643      & 0,58597        \\
10\%               & 0,88742        & 0,45455       & 0,83165        & 0,42308       & 0,84306         & 0,51613        & 0,98643      & 0,58597        \\
11\%               & 0,88742        & 0,45455       & 0,85306        & 0,45714       & 0,85693         & 0,53333        & 0,98643      & 0,58597        \\
12\%               & 0,88616        & 0,45455       & 0,85956        & 0,44118       & 0,86018         & 0,50847        & 0,98662      & 0,58597        \\
13\%               & 0,88490        & 0,45455       & 0,88216        & 0,43750       & 0,87305         & 0,49123        & 0,98688      & 0,58597        \\
14\%               & 0,88289        & 0,45455       & 0,90265        & 0,47368       & 0,88450         & 0,51064        & 0,98884      & 0,58597        \\
15\%               & 0,88289        & 0,45455       & 0,90265        & 0,47368       & 0,88450         & 0,51064        & 0,98884      & 0,58597        \\
16\%               & 0,88163        & 0,44000       & 0,91327        & 0,47368       & 0,88980         & 0,47826        & 0,98908      & 0,58597        \\
17\%               & 0,87197        & 0,44000       & 0,91538        & 0,47368       & 0,88580         & 0,48889        & 0,98912      & 0,58597        \\
18\%               & 0,87197        & 0,44000       & 0,91538        & 0,47368       & 0,88580         & 0,48889        & 0,98912      & 0,58597        \\
19\%               & 0,87086        & 0,44000       & 0,93165        & 0,47368       & 0,89476         & 0,50000        & 0,98946      & 0,58597        \\
20\%               & 0,86130        & 0,35484       & 0,93645        & 0,44444       & 0,89218         & 0,46809        & 0,99000      & 0,58597        \\
21\%               & 0,86130        & 0,35484       & 0,93645        & 0,44444       & 0,89218         & 0,46809        & 0,99000      & 0,58597        \\
22\%               & 0,86130        & 0,35484       & 0,95054        & 0,61111       & 0,89886         & 0,46809        & 0,99000      & 0,58597        \\
23\%               & 0,86130        & 0,35484       & 0,95054        & 0,61111       & 0,89886         & 0,46809        & 0,99000      & 0,58597        \\
24\%               & 0,85805        & 0,32000       & 0,95718        & 0,64286       & 0,90046         & 0,43243        & 0,99165      & 0,58597        \\
25\%               & 0,85805        & 0,32000       & 0,95718        & 0,64286       & 0,90046         & 0,43243        & 0,99165      & 0,58597       
\end{tabular}
\end{table*}

\begin{figure}[ht]
  \centering
  \includegraphics[width=\linewidth]{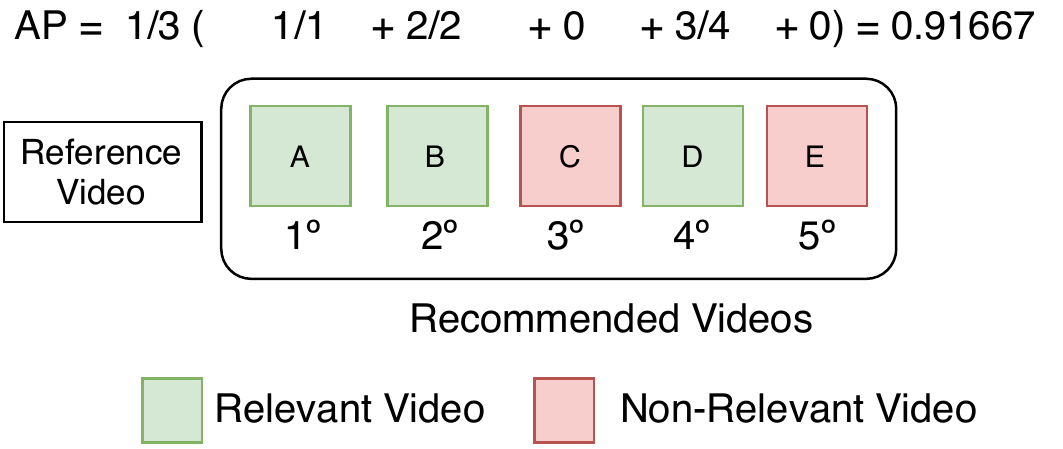}
  \caption{Example of how the Average Precision~(AP) is computed for a reference video and its recommended videos.}
  \label{fig:ap_example}
\end{figure}

To evaluate our ranking, for each video we compute the Average Precision~(AP), that evaluates how well a ranking of recommendations is based on each element's relevancy. 
This metric penalizes more a ranking if a non-relevant element is recommended in the first positions than if it was in the last ones.
Let $P_k$ be the precision of the first $k$ elements of a ranking, which is the percentage of videos that are relevant in the sub-ranking that starts at position $1$ and ends at position $k$.
Let $\alpha_k$ denote the relevancy of the video in position $k$, where $\alpha_k = 1$ if the video is relevant, and $0$ otherwise.
The AP of a given ranking is defined as follows
\begin{equation}
  \label{equation:average_precision}
  AP = \frac{1}{GTP}\sum_{k=1}^{n}{P_k~\cdot~\alpha_k}
\end{equation}
where GTP refers to the total number of ground truth positives in the ranking, which is the total number of videos that are considered relevant in a ranking. Fig.~\ref{fig:ap_example} shows an example of how the AP is computed for a given ranking. In this case, the $GTP=3$ because the total number of relevant videos in the ranking is 3~(videos A, B and D).

In order to prevent outliers from having much influence in the recommendation (e.g. a person that is not a lecturer -- and not relevant to the video -- and appears for a short amount of time), we experimented different thresholds of presence intervals in a video for a person to be considered as ``present'' when computing the score for the ranking. 
In this way, a $p_{l,v}$ lesser than the threshold is considered as $0$.
Besides the Average Precision, we also compute the mean and minimum values of the recall~(MeanR and MinR), precision~(MeanP and MinP), and F1-Score~(MeanF1 and Min F1) for the recommendation generated for each of the videos, without considering the positioning of these videos in the rankings.
The recall refers to the percentage of relevant videos that are present in the ranking.
The F1-score represents an overall performance metric based on the  harmonic mean of the precision and recall and is defined as follows.
\begin{equation}
    \label{equation:f1}
    F1 = \frac{2 \cdot P \cdot R}{P + R}
\end{equation}
Table \ref{tab:results} shows the thresholds used, the values of recall, precision, F1-score, and the mean and minimum Average Precision (mAP and MinAP),

One can observe from Table \ref{tab:results} that the precision clearly increased with the use of the threshold.
Different from the precision, the recall decreased with the increase of the threshold. It means that with a greater threshold more videos that should be recommended were not chosen by our method.
It is important to notice that these two metrics~(precision and recall) do not consider the ordering of the recommendations.
Different from them, the Mean Average Precision~(mAP) has high values for all thresholds, specially because the score for computing the ranking takes into consideration the percentage of time that a person appears in the reference and recommended videos.
Then, we can conclude that our proposed approach for ordering the recommended videos tends to recommend more suitable videos first with a high mAP$\approx0.99$. Moreover, despite the precision of the \emph{Clustering Step} shown in Table \ref{TabelaResultadosVideoAnnotationTool}, our method for ranking was robust to outliers and was able to correctly recommend and rank relevant videos.

\section{Final Remarks}
\label{sec:remarks}

In this paper, we present a method for educational video recommendation using deep-face-features of lecturers. More precisely, we use an unsupervised clustering-based method and an heuristic for ranking.
It takes advantage of face detection mechanisms to perform educational video recommendation based on the lecturers' presence.
Besides the face detection, we also perform face clustering of the lecturers in each video, and, given these clusters, we extract their centroids to perform another clustering step that creates a relationship of videos that share the presence of the same lecturers.
Finally, we rank the recommended videos based on the amount of time each lecturer is present.
It is worth mentioning that our method is completely automatic and does not require any information of the video files in advance.  
Moreover, our approach does not need to know or store the identity of the lecturers for performing recommendation, preserving their privacy.

\begin{figure}[!ht]
  \centering
  \includegraphics[width=\linewidth]{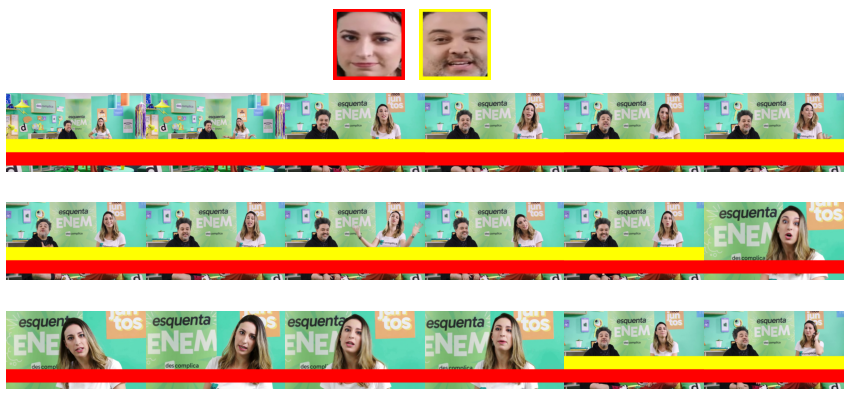}
  \caption{Educational video timeline tagged by the lecturers presence. Notice that the frames with the lecturer on the left are tagged in red, while the frames with the lecturer on the right are tagged in yellow. This timeline is resulted from the \emph{Video Representation} phase, described in Section \ref{subsec:video_representation}.}
  \label{fig:video_timeline}
\end{figure}

A collateral contribution of our paper is video segmentation the by lecturer.
As illustrated in Fig.~\ref{fig:video_timeline}, we can create a timeline based on lecturers' presence, which can be used to help students in finding moments where specific lecturers are present.
With this segmentation, we could recommend specific parts of the video to the student.

The main limitation of our work is that we can only recommend videos in which the lecturers are visually present. As future work, we intend to use a hybrid recommendation approach, that combines both textual and audiovisual information from the video to create clusters. Video summarization is also a technique that can be explored to enhance video content searching and selection.


\bibliographystyle{IEEEtran}
\bibliography{IEEEabrv, refs}
\end{document}